# Avalanche-like lithium intercalation and intraparticle correlations in graphite

Jiho Han[1,2,4], George S. Phillips[2,4], Alice J. Merryweather[1,2,4,5], Juhwan Lim[1,4], Christoph Schnedermann[1,4,5], Robert L. Jack[2,3], Clare P. Grey*[2,4], Akshay Rao*[1,4]

1. Cavendish Laboratory, University of Cambridge, J.J. Thomson Avenue, CB3 0US, Cambridge, UK. E-mail: ar525@cam.ac.uk
2. Yusuf Hamied Department of Chemistry, University of Cambridge, Lensfield Road, CB2 1EW, Cambridge, UK. E-mail: cpg27@cam.ac.uk
3. DAMTP, Centre for Mathematical Sciences, University of Cambridge, Wilberforce Road, Cambridge CB3 0WA, United Kingdom
4. The Faraday Institution, Quad One, Harwell Science and Innovation Campus, OX11 0RA, Didcot, UK
5. Illumion Ltd, Maxwell Centre, J.J. Thomson Avenue, CB3 0HE, Cambridge, UK

*email:
Clare P. Grey: cpg27@cam.ac.uk
Akshay Rao: ar525@cam.ac.uk

## Abstract

Graphite is the most widely used anode material in lithium-ion batteries with over 98% market share. However, despite its first application over 30 years ago[1], the lithium insertion processes and associated dynamics in graphite remain poorly understood, especially for the dilute stages. A fundamental understanding of how the symmetry-breaking phase transitions occur pseudo-continuously, in the disordered graphites used in batteries, and under operating conditions is still lacking. Here, we provide a unified picture of ion intercalation dynamics during the dilute stages of graphite intercalation, using *operando* optical scattering microscopy combined with random field Ising modelling. We show that during the dilute stages, single graphite particle undergoes rapid, localised avalanche-like (de)intercalation events, leading to micron-sized regions (de)intercalating within seconds. These avalanches are reminiscent of phase transition behaviour seen in disordered materials such as martensitic transformations[2,3], Barkhausen noise[4] and ferroelectric and ferroelastic materials[5,6]. They are associated with step changes in the order parameter, where the system transitions from one phase to another under an applied driving force, by jumping from one metastable state to another. Here, using a modified random field Ising model, we relate these avalanches to static disorder in graphite, which disrupts ion filling dynamics, leading to "pseudo-continuous" transitions between stages, accounting for the experimental electrochemistry profile as well as the temperature dependent avalanche dynamics. Finally, we develop a methodology to spatio-temporally analyse avalanches between intraparticle regions, revealing spatially heterogeneous connectivity and temporal patterns between regions during the dilute stages. Our work highlights the role of local and static disorder in eliciting unexpected phase transition behaviour, and provides new tools and concepts for studying layered battery materials.

## Introduction

Since its first application in 1980s[1,7,8], graphite has become the anode material of choice in nearly all lithium-ion batteries. Owing to its layered structure, lithium intercalation into bulk graphite proceeds via a series of distinct transitions between stages, which are labelled according to their stage number '$N$' denoting that every $N$-th inter-graphene gallery contains lithium. Lithiation begins with a dilute stage 1'L, where every layer contains few, unordered lithium ions, and then proceeds via transitions through stages 4L→ 3L → 2L → 2 → 1. The suffix 'L' indicates a dilute phase with no lithium order in the filled gallery, and stage 1 corresponds to the well-known, closely packed lithiated configuration of graphite, $LiC_6$.

Despite its ubiquity, previous studies on graphite have resulted in contradictory reports on phase composition during *operando* cycling, particularly in the low lithium-concentration "dilute stages"[9–11]. Phase transitions in battery materials are often categorized as a solid-solution[12,13] or phase separating[14–17], where lithiation in graphite has been reported to have both first-order biphasic transitions and "pseudo" solid-solutions[9,18,19]. Such dynamics are usually studied electrochemically and with advanced *operando* techniques such as X-ray, neutron or electron diffraction. However, the dilute stages in graphite are particularly difficult to study due to low atomic weights, rapid lithium diffusion[20,21], metastable phase formation and particle-level heterogeneity[22,23], which necessitate high spatial and temporal resolution, and makes them more prone to beam damage.

Here, we use the recently developed method of charge photometry, an *operando* optical scattering microscopy, to investigate the sub-second dynamics of lithium intercalation in an individual active particle of graphite during the dilute stages as a function of temperature. Empowered by this technique, we show that (de)lithiation of the dilute stages in graphite proceeds via a series of step-like events, avalanches, spatially localised within an intraparticle region. This is fundamentally different from our previous optical microscopy studies where we observed first-order biphasic behaviour[24] or solid-solution diffusion[25,26] within cathode materials. The events in graphite cluster around key peaks in differential capacity plots, and resemble the "avalanches" observed during phase transitions in materials possessing static (quenched) disorder[27]. We develop an anisotropic random field Ising model (RFIM) to describe the energy landscape of lithium intercalation in a disordered lattice, which is capable of qualitatively modelling the experimentally observed temperature-dependent avalanche trends and electrochemical hysteresis curves, as well as reproducing the "pseudo" solid-solution 4L-3L transition described in literature.

Our results reveal that the dilute stages transition through shallow, metastable states via small and rapid lithium flux in a local region, where each avalanche corresponds to a layer filling of a single or few correlated domains. This results in a lack of pure "stages" and highly mixed filling periodicity, which corresponds well to the broad reflections seen in previous diffraction studies and attributed to disorder and phase coexistence throughout the dilute stages[9,10,28]. Finally, using the cross correlation function as an effective way of analysing the stochastic filling events, we demonstrate that the seemingly random sequence of avalanches revealed in our optical experiments possess definite spatio-temporal correlations patterns, related to the regions and disorder identified in the electron backscatter diffraction (EBSD) map.

**Optical signatures of graphite lithiation and avalanche-like jumps**

To study the ion dynamics of graphite on a single particle level, an optically-accessible half-cell (coin cell format) with a self-standing graphite electrode was prepared and imaged using an optical scattering microscope[24] equipped with a temperature-controlled cell holder (Fig. 1a). An optical image of a representative particle is shown in Fig. 1b (upper panel) alongside its scanning electron microscopy (SEM) image taken after the experiment (lower panel), highlighting the single graphite particle embedded in a carbon electrode. While the studied region of the particle (highlighted via a red dotted line, Fig. 1b top) was confirmed to correspond to the graphite basal plane, high-resolution SEM and electron backscatter diffraction (EBSD) mapping showed that it was comprised of multiple sub-domains (Supplementary Fig. S2a-d).

During the formation cycle, the electrode was initially cycled against a lithium metal counter electrode to 5 mV vs Li/Li$^+$, to achieve full lithiation (LiC$_6$) with a constant current of 7 μA (corresponding to a C-rate of C/20). The cell achieved a discharge capacity of 402 mAh g$^{-1}$, slightly higher than the theoretical capacity of 372 mAh g$^{-1}$, due to additional irreversible capacity associated with the formation of a solid electrolyte interphase (SEI) layer on graphite and the large amount of high-surface-area conductive carbon present in the electrode. The voltage curve, shown in Fig. 1c (top), exhibits the familiar staging behaviour of graphite, with dilute stages occurring from 205 - 115 mV, followed by the stage 2L → 2 transition at a voltage plateau of ~112 mV and the stage 2 → 1 voltage plateau at ~77 mV.

Fig. 1c (bottom) shows the corresponding optical scattering intensity obtained during this cycle for the highlighted flake (Fig. 1b, top). The mean intensity initially drops with lithiation in the dilute stages, which is followed by a significant increase in intensity, at the stage 2L → 2 and stage 2→ 1 transitions, both of which are biphasic, first-order phase transitions according to diffraction studies[9]. The optical response of graphite during lithiation has been well studied[29], and can be explained in terms of increasing electron carrier density with lithiation, resulting in a plasma frequency ($\omega_p$) blue-shift and increased metallic reflection. Importantly, the optical intensity is a direct probe for the local lithiation state as discussed previously [25,26] and here, during the dilute stages, the intensity decreases monotonically during lithiation, opposite to that seen for the dense stages. The penetration depth is at most ~100 layers and within the particle depth (see Supplementary discussions).

To gain further insight, we continued to cycle the above cell over a smaller voltage window at the same cycling rate from 112 to 240 mV, as shown in Fig. 1d (upper panel), where transitions between the dilute stages are expected. The optical scattering intensity in this voltage region of the graphite particle (Fig. 1d lower panel) exhibits features which compare well to expected transitions from the voltage profile. A sharp drop in intensity occurs during the voltage plateau around ~200 mV vs Li/Li⁺ where the first-order 1'L→ 4L transition is expected. Subsequently, we observed a seemingly continuous drop in optical intensity during the pseudo solid-solution transition 4L→3L where the voltage gradually drops to ~120 mV. At this point, a final drop in optical intensity occurred, in line with the expected 3L→ 2L transition. Crucially, when imaging at sufficiently high frame rates (2 Hz or 0.0007% $Li_x$ per frame), these seemingly-continuous intensity changes were revealed to consist of a sequence of many small, sudden, discrete jumps, such as shown Fig. 1f. These step-like features in the optical intensity were observed during both de/lithiation and also across cells with different graphite types (Supplementary Fig. S3) when cycled at the same rate (C/20). Furthermore, these step-like change could be attributed to a process involving a distinct, localised region of the graphite particle, as illustrated in Fig. 1e using differential image analysis, which highlights the changes in intensity over 30 seconds. Fig. 1f demonstrates the intensity profile of some of these local particle regions (Fig. 1f inset regions A, B, C), further demonstrating that the step-like lithiation behaviour is observed across the graphite particle.  It is important to stress that these step-features do not occur during the dense stage 2 – 1 transition, where we observe an intercalation wave like mechanism[24,30] with nucleation and growth of a single new phase across the entire particle consistent with a first order biphasic transition (Extended Data Fig. 1).

**Step-events correlated with electrochemical features**

To further understand the optical intensity behaviour during the dilute stages, the optical response was correlated with the cell voltage in Fig. 2a for different regions of the particle (B, C, D, Fig. 2d) as a function of different cell temperatures. During lithiation and across all temperatures, we observed a large sharp intensity drop at ~200 mV, followed by a more gradual decrease (still containing many small discrete steps), with another subsequent steeper drop at ~120 mV. The sharp-gradual-sharp intensity trend (with respective regimes labelled (i), (ii), (iii) in Fig. 2a upper-right) also occurs during delithiation; the voltage hysteresis between (de)lithiation cycles increases at lower temperatures but the small step-like intensity changes appear throughout all datasets in all three regimes (i)-(iii).

To quantify the intensity steps further, we employed a step detection algorithm to analyse the intensity traces from the individual graphite regions (see Supplementary Methods). This approach allowed us to build up statistics of these events as a function of temperature, cycle, voltage and region-dependence. Each detected step event is plotted as a circle against the voltage at which it occurred (Fig. 2b upper panel – delithiation, lower panel – lithiation). The intensity change during the event (i.e. magnitude of the step fall or rise) is then shown as the size of the plotted circle. For simplicity, we have limited ourselves to events in region B of the particle in this plot. The distribution of the events is then correlated with a differential capacity analysis (dQ/dV) of the cell electrochemistry, which highlights underlying phase transitions or processes. Fig. 2b (middle panel) shows the dQ/dV plots at different temperatures, with prominent peaks marked with numbers. The first peak to occur during lithiation at 197 - 202 mV (1) is attributed to the 1'L → 4L transition. There is a very small peak occurring around 162 mV (2a) that becomes more pronounced at higher temperatures (see Supplementary Fig. S4a), then a larger peak near 140 mV (2b). An additional peak (3) is visible as a shoulder of the large truncated peak (assigned to the onset of the 2L→2 transition). During delithiation, similar trends are observed. However, the smaller peaks (2a) and (2b) are more prominent and are present at all temperatures. Based on this analysis, the two sharp drops in optical intensity shown in Fig. 2a-(i),(iii) are attributed to the differential capacity peaks (1) and (3), respectively, while peaks (2a,b) occur during the more gradual intensity changes (Fig. 2a-(ii)) (Supplementary Fig. S4b). Furthermore, the step-like events detected in Fig. 2a (upper, lower panels) cluster around the dQ/dV peak positions, with the events becoming more tightly clustered at higher temperature. This is closely related to the dQ/dV peak shape, which sharpens at higher temperature.

The same analysis can be extended to other regions of the particle and a selection is shown in Fig. 2c for the 45 °C delithiation cycle. Regions A, B and C show consistent trends as discussed but region D show events which are more widely distributed, even at 45 °C. This was true generally, where some regions consistently showed a lag and broad temporal distribution of events at all temperatures (regions D and E, Supplementary Fig. S5). Critically, grain reference orientation deviation (GROD) map derived from the EBSD measurement show in Supplementary Fig. S1e that these regions may be associated with localised disorder, with areas associated with region D and E showing a higher change in GROD with distance. This also affects intraparticle connectivity, which is discussed further below.

**Phase transition behaviour of a disordered system with ramped potential**

The step-like intensity features of the individual regions revealed through optical imaging in the dilute stages are highly reminiscent of "avalanches", as observed in martensitic transformations, Barkhausen noise and ferroelectric and ferroelastic avalanches. In those systems avalanches represent changes in the order parameter, due to jumps from one metastable state to another, as the system transition from one phase to another under an applied driving force (such as external field or potential)[31] and are often associated with static (quenched) disorder, which graphite is known to be affected[32,33]. To test this analogy, we developed a 2D random field Ising model (RFIM) to explore the energy landscape of lithium staging behaviour in graphite, where we simulate a graphite lattice with discrete lithium occupation with attractive first-neighbour intralayer coupling, up to third-nearest neighbour repulsive interlayer coupling and on-site static random potential representing idealised disorder (Extended Data Fig. 2a). The lattice filling behaviour is ruled by dimensionless disorder strength ($\tilde{h}$), temperature ($\tilde{T}$) and applied potential (field) ($\tilde{\mu}$), which are measured relative to intralayer interactions. A full description of the model is presented in the Supplementary Methods. This model is particularly applicable as dilute stages are associated with fast diffusion and disordered, heterogeneous phases with complex transition dynamics.

In Fig 3a, we simulate a lattice with medium disorder ($\tilde{h} = 0.4$) at low temperature ($\tilde{T} = 0.125$) and plot its mean fill density ($\phi$) with time. The filling curve (Fig. 3a upper panel) shows a sharp transition from empty to stage 4-like filling ($\phi \sim 0.21$) (i), then a sloping transition from quasi-stage 4 to quasi-stage 3 ($\phi$~0.31) (ii), and finally a sharp transition from quasi-stage 3 to quasi-stage 2 ($\phi$~0.45) (iii). We emphasize that pure staging no longer exists at any point in this model, demonstrated by the broad, ill-defined peaks in the c-axis Fourier transform of the Li-ion positions (Fig. 3a lower panel). To further highlight this aspect, it is instructive to evaluate the c-axis distance between occupied sites in the lattice. For a stage-n phase, we expect these distances to be equal to n, except for domain interfaces and random thermal fluctuations. However, the lattice picture (Fig. 3b) shows a lack of long-range order and a mixing of distances between filled sites (intergallery distance). The proportion of n-separated intergallery distances found for the lattice at different times is shown in Figure 3c. At every time point there exists always a mixture of distances and critically, we still observe the same trend seen in the experimental mean intensity, comprising of a sharp transition to stage 4-like filling (Fig. 3a point (i)), a continuous-like transition to stage 3-like filling (Fig. 3a point (ii)), followed by another sharp transition to stage 2-like filling (Fig. 3a point (iii)). We label the critical points (i)-(iii), stages 4M, 3M and 2M respectively, with M implying a highly mixed stage character to distinguish them from true stages with long range order.

At zero disorder ($\tilde{h} = 0$) and higher temperature ($\tilde{T} = 0.5$), the model can exhibit true stages without mixing disorder. This is illustrated in Supplementary Fig. S6, where long range correlation results in system wide transitions between empty, stage 4, 3 and 2. This demonstrates that static disorder can break up interlayer correlation, leading to local domains, avalanches and metastability. A schematic of model behaviour with varying temperature and disorder is shown in Extended Data Fig. 2b, where the intermediate temperature and disorder regime (marked with star) is relevant to our experiment. Due to decaying interlayer repulsion, the system is most affected by disorder for higher order stages, resulting in a rounded, continuous-like transition between stage 4M and 3M. The significance of static disorder on phase transition behaviour is well studied in standard RFIM[34,35], where long range correlations, which allow a sharp, system-wide first-order transition, is broken at higher temperature and by disorder, thereby making the transition more rounded and continuous. Near the critical line, disorder allows the formation of finite metastable domains, and avalanches appear as each local domain fills (Extended Data Fig. 2c).

In our lattice model, an avalanche often corresponds to a single layer (or a few correlated layers) filling in a local domain, as demonstrated in Supplementary Fig. S7. We note that our model is an idealised representation of disorder such as stacking faults and turbostratic disorder in graphite. In reality, as lithium diffuses into graphite, a layer opening barrier, grain boundary diffusion barrier and shear stresses from AA to AB layer stacking[9] will add to the complexity of disorder. It is interesting to note that avalanches still appear with a more crystalline graphite sample (Kish graphite, Supplementary Fig. S3c,d), indicating that other sources of disorder may contribute to avalanches in the graphite dilute stages.

**Avalanche distribution and lattice filling dynamics**

Importantly, our model reproduces many other features of the experimental data. For example, the filling/emptying hysteresis is temperature dependent, with increasing hysteresis at lower temperature (Fig. 4a) much like the experimental intensity curves (Fig. 4b). Furthermore, avalanche events can be detected from the simulated filling curves, where avalanches occur more frequently at higher temperature (Fig. 4c). This is expected from previous studies on standard RFIM[36] where at higher temperatures, (1) large (high $\Delta\phi$) avalanches occur less frequently and (2) smaller avalanches occur more frequently, due to thermal fluctuations exciting smaller avalanches and consequently prevent the build-up required for large avalanches to occur. The experimental avalanche statistics are dependent on the region sampled and the voltage window, and here obtained from region A, B and C in the 1'L-4L voltage window in cycle 1 and 2. The total number of detected events are shown in Fig. 4d, demonstrating that avalanches become more frequent at higher temperature, as is especially clear for delithiation.

Interestingly, thermal fluctuation at higher temperatures also cause reverse avalanches, here defined as a step change in the opposite direction of the driving force. These are shown in the simulations in Fig. 4e, where reverse avalanches appear at higher temperatures. We also experimentally detect these reverse avalanches, where a local region seemingly delithiates during a lithiation cycle and vice versa, as demonstrated in Supplementary Fig. S8. These are generally small and become far more frequent at higher temperature (Fig. 4f), supporting the idea that the avalanches are thermally activated jumps between shallow metastable states. Critically, it implies that we are in a regime where thermal fluctuation, the disorder strength and long-range coupling energies are all relevant, and of similar energy (~kT).

The avalanche size distributions are shown as a violin plot in Supplementary Fig. S9b-c where avalanche size is defined the change in mean intensity. While the simulation trends are as expected, with larger avalanches cut-off and smaller (and reverse) avalanches becoming more common with higher temperature, the experimental trend is inconsistent, where large magnitude avalanches remain common at higher temperature. However note that the step-detection algorithm is prone to detecting multiple step events as a single large event as the time separation between steps decreases.

Therefore, at higher temperature as the events bunch more in time (see Fig. 2b), clusters of smaller events may frequently be detected as a single large avalanche, resulting in deviation from the trend.

An interesting consequence to static (quenched) disorder is that the model runs are sensitive to the structure of this randomness. As shown in Extended Data Fig. 3a, when the disorder structure is kept the same, the model returns similar results over many runs, as seen by the similarity of the set of green or red traces, respectively. The model supports our experimental observation presented in Extended Data Fig. 3b, whereby each spatial region of the graphite particle displays similar curves with avalanches at similar positions over different cycles, providing evidence for a unique, structural disorder in each region.

**Spatio-temporal correlation of avalanche-events**

Since our dataset contains a series of "spikes" in the differential intensity time series, this opens up analysis routines frequently exploited in time-domain correlation spectroscopy. Here, we calculate the pixel-pixel pair cross correlation function for the differential time series images. The full methodology for this section can be found in Supplementary methods. This general method enables us to investigate our dataset for spatio-temporal correlation without relying on manual region selection or avalanche detection algorithms, and we can extract a cross-correlation matrix showing the relationship between any two pixels at a given time-lag.

First, we explored the correlation matrix where off-diagonal elements signify pixels which undergo a concerted intensity change within the defined time period of ±5 s to which spectral co-clustering was appto identify the blocks of pixels that are correlated (Fig 5a (i)). When these blocks are mapped back to the particle, spatial regions emerged (Fig 5a (ii)) which well reproduce the manually selected particle regions presented above as well as the regions identified in the EBSD map (Supplementary Fig. S2c-e). We repeated this analysis for different temperatures, (de)lithiation cycles and lag-times, and the regions remained similar throughout (Supplementary Fig. S10). The clusters identified in Fig. 5a have a strong preference to undergo (de)lithiation together, however the presence of off block-diagonal elements highlight the existence of strong inter-region correlations.

To explore these correlations, the data set was coarse-grained into distinct regions, which were used to re-calculate the correlation function. This allowed us to identify the regions with most inter-region connectivity. This is shown in Fig. 5b, where these coarse-grained regions are coloured according to most and least inter-region correlations ($-5$ s $\leq \tau \leq 5$ s). The results are highly consistent with our previous analysis, where the least connected regions (4,5 in Fig. 5b) are the same as those found to have step-events which are broadly distributed around dQ/dV peaks (region D in Fig. 2c, region E in Supplementary Fig. S6).

To explore specific inter-region relationships, the correlation matrix is summed over a positive time-lag ($3 \text{ s} \geq \tau > 0$), which signifies that an event in one region is followed by an event at another region within the time window. The resultant correlation matrix (Fig 5c(i)) is asymmetric due to one region preferring to precede the other. The results of this analysis in a directed graph network, with the directionality and weight of the edge corresponding to the time-ordered difference of the corresponding off-axis elements, as represented in Fig 5c(ii). The arrow then indicates likely predecessor-successor relationship between the regions. We then extend this analysis to all cycles and temperatures, where inter-region correlations are collected at different lag times and kept only if they occur in the majority of datasets. In Fig. 5d, these relationships are shown on the particle which demonstrates that at small time-lag ($\tau \sim +1$ s), well-connected, adjacent regions become correlated whereas these correlations become longer range at larger time-lags ($\tau \sim +5$ s). Interestingly, the correlations significantly differ between lithiation and delithiation. Notably, the short range, short time-lag correlations often flip between de/lithiation. This implies that the correlations are not simply due to the directionality of phase fronts as seen in the dense stages (Extended Data Figure 1) or differences in local overpotential. Rather, each avalanche is a transition to a reoccurring, spatially localised metastable configuration, which may account for the asymmetry during (de)lithiation.

**Conclusion**

This study highlights unexpected intercalation behaviour which is at odds with the current battery literature, and yet is well understood within the context of driven systems with quenched disorder. In these systems lithiation may proceed via jumps between metastable filled states, where each jump corresponds to a rapid but small increase in concentration by filling a single, or a few-correlated layers simultaneously in a localised region of the particle. Macroscopically, the ensemble behaviour may resemble a solid-solution like transition, yet at a single particle, the behaviour is markedly different to both phase-separating (as seen for the dense stages) and solid solution systems.

Notably, the avalanches or rapid filling of the layers occurs within sub-regions of the larger particle in the earlier dilute stages, while the stage 2-1 biphasic reactions occur as a "wave" progressing across the whole imaged particle. The stage 2L-2 transition behaviour is intermediate between the two, where the layers are weakly coupled and fill layer-by-layer in quick succession (See Supplementary Figure S11). This suggests that the stronger interlayer coupling in the dense stages results in a more traditional nucleation and growth mechanism; when these interlayer couplings are much weaker in the dilute stages, the layers act semi-independently due to correlation breaking disorder and the defects/activation barrier associated with filling an individual layer dominates the intercalation process. This is supported by the modelling, which suggests that pure dilute stages may not exist, even at a single particle level, and it is more appropriate in a more disordered graphite to view the process as involving successive filling of layers, so that increasingly stage 4L, 3L and then 2L stacking dominates. The effectiveness of our simple model motivates the incorporation of disorder in more realistic, continuum models.

The dynamics shown here redefine our understanding of phase transitions in realistic, disordered and layered battery materials. And our results demonstrate the power of operando single-particle techniques, and in particular optical scattering microscopy, to advance our understanding of battery materials for future sustainable applications. We anticipate that the methodologies, both experimental, analytical and modelling, presented here will drive the re-examination of disorder and discovery of new, battery materials.

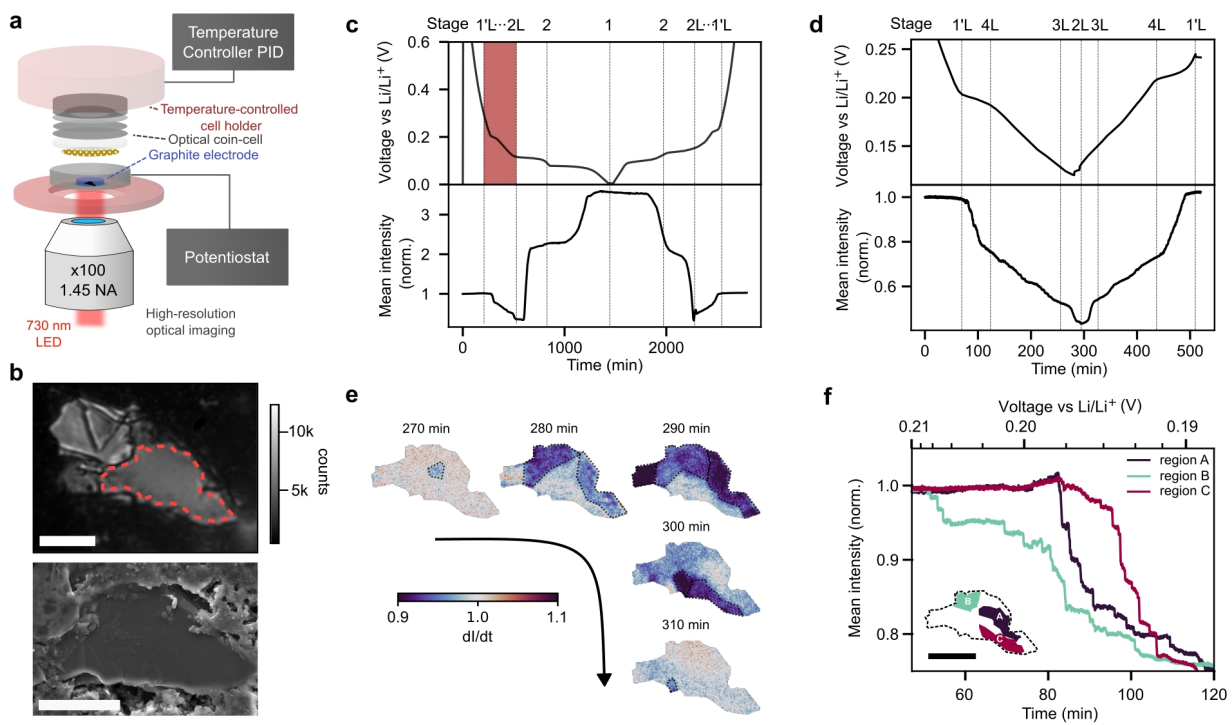

**Figure 1 | Experimental set-up and observation of spatially local discrete intensity changes. a,** Battery coin cell and optical microscopy set up. An optically-accessible coin cell (full schematic shown in

Supplementary Fig. S1) with a self-standing graphite working electrode and lithium metal counter/reference electrode. The cell is connected to a temperature-controlled cell holder and potentiostat. **b,** Optical scattering image of the active graphite particle (top). The dotted red region indicates the area of the particle analysed in this work. Ex-situ SEM image of the same graphite particle (bottom). Scale bars 5 μm. **c,** Voltage plots during one full cycle between 1 V and 5 mV at C/20 (7 μA) rate at 25°C in black (upper panel). The red highlighted region indicates the dilute stages regime studied here. Particle-averaged optical scattering intensity during cycling (lower panel), from the red-dotted region shown in **b**. Norm. indicates normalised to initial intensity (in **c, d, f**). Vertical dashed lines (in **c** and **d**) indicate the phase (stage number) expected at the given time/voltage. **d,** Voltage plot during cycling over a narrower voltage window at C/20 (7 μA) rate at 25°C (upper panel). Particle averaged optical scattering intensity during cycling (lower panel). **e,** Sequential differential contrast images of the graphite particle during the dilute stage transition (stage 1'L to 4L). Images are obtained by dividing respective frames by another taken 30 s earlier. Blue (red) indicate an instantaneous decrease (increase) in optical scattering intensity. Images indicate fast and localised drops in intensities in distinct regions of the graphite particle. **f,** Mean optical intensity during stage 1'L to 4L transition, normalised to the start of cycle for distinct graphite regions. Regions of the particle shown in bottom left for reference, scale bar 5 μm. Upper x axis indicate the cell voltage interpolated from the time.

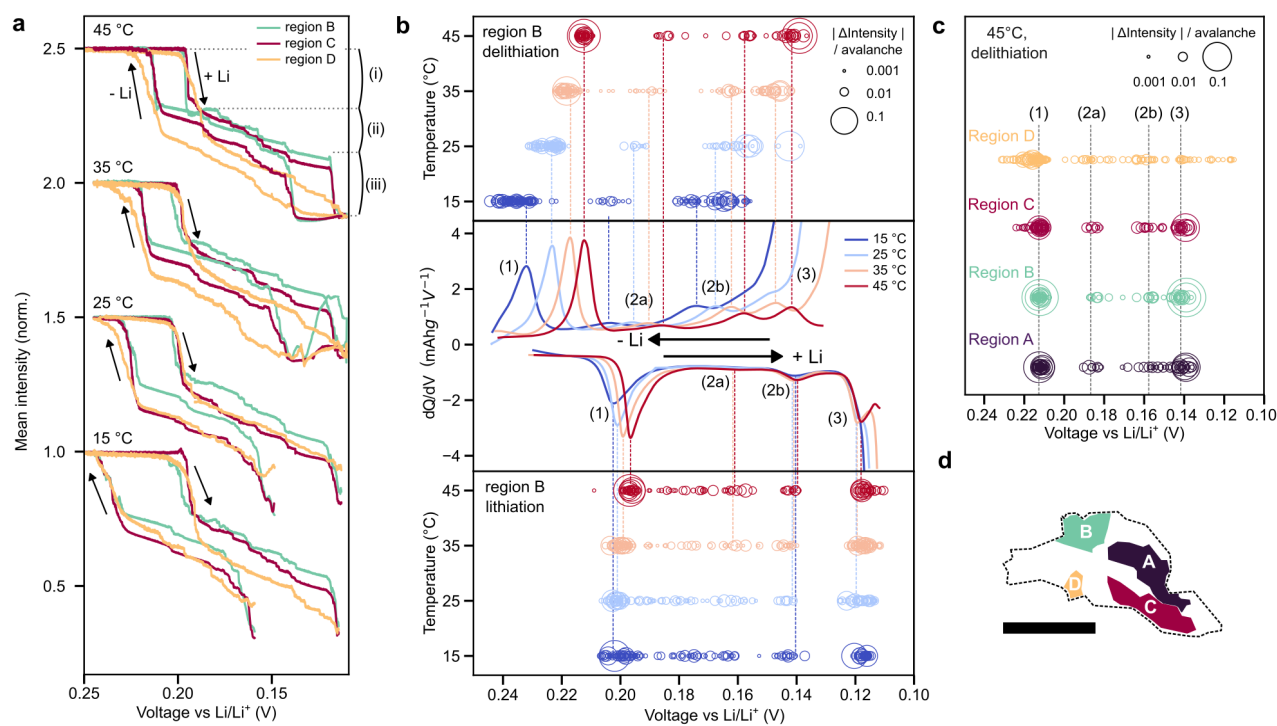

**Figure 2 | Optical step-changes correlated with cell voltage. a,** Mean optical intensity curves from different particle regions, versus voltage, normalised to the start of lithiation or end of delithiation. Lithiation and delithiation curves are both shown, direction denoted by black arrows. Only regions B, C, and D in the particle are shown for simplicity. (Where appropriate) the curves have been cut before the onset of the stage 2L to 2 transition for clarity. Three regimes in the intensity trends have been marked (i) – (iii). **b,** Differential capacity (dQ/dV) curves from the charge/discharge data at different temperatures (middle panel). The first

cycle at each temperature is shown. Four processes labelled (1), (2a), (2b), (3) can be distinguished. The onset of a peak corresponding to the dense transition (2L to 2) is seen in these plots at temperatures of 35 °C and below. Black arrows denote the direction of lithiation vs delithiation. Scatter plot of the avalanches that occur during delithiation (upper panel) or lithiation (lower panel), detected from the normalised mean optical intensities in region B, in cycles 1 and 2, as a function of temperature. Each circle represents one avalanche event, with its x-position denoting the voltage at which it occurs and the size denoting the change in intensity. **c,** Scatter plot of the detected avalanches in different regions of the particle, during 45°C delithiation (cycle 1 and 2). The x-position denotes the voltage at which the avalanche occurs and the size denotes the intensity change. Voltages corresponding to processes (1), (2a), (2b), (3) as found in **b** are marked with the dotted line. **d,** Intraparticle regions, labelled A-D, used for mean intensity calculations. Sscale bar 5 μm.

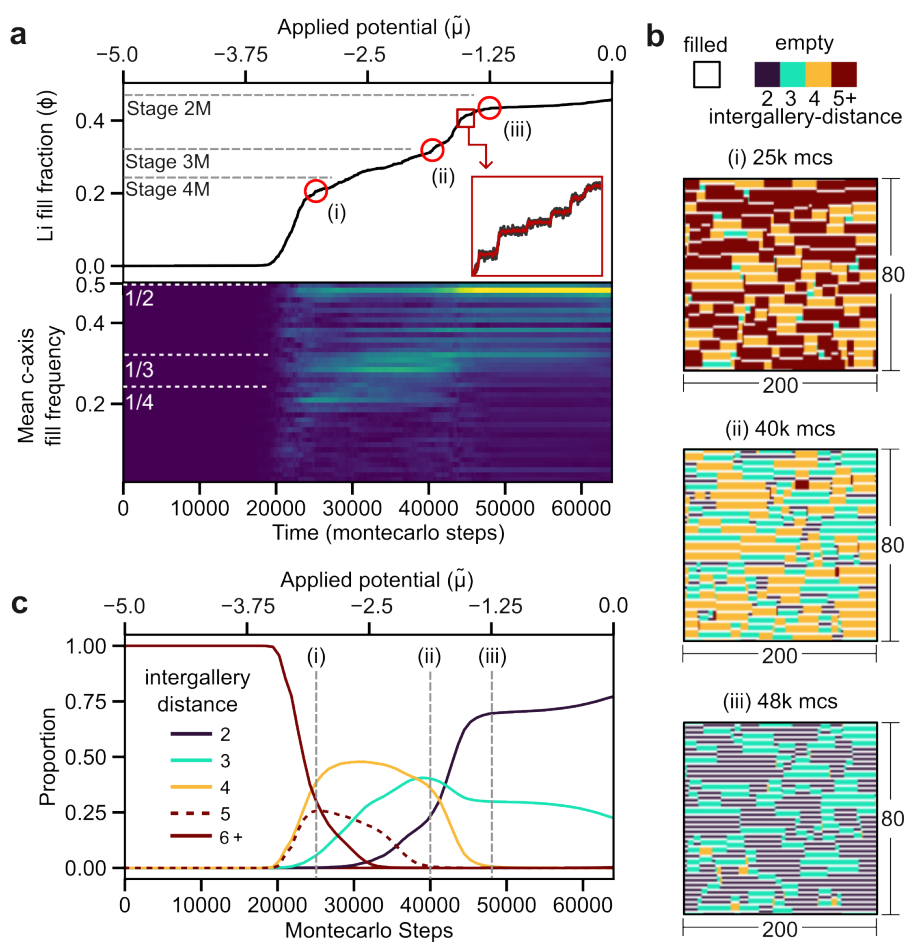

**Figure 3 | Disordered lattice model (random field Ising-like) filling behaviour . a,** Mean lattice fill fraction ($\phi$) vs Monte Carlo steps (mcs) during a single ramped potential simulation (upper panel), with disorder ($\tilde{h} = 0.4$), and temperature ($\tilde{T} = 0.125$), with total mcs = 80k. Red circles denote time points (i)-(iii) at which the lattice pictures in **(b)** are taken from. Red square inset highlights a portion of the filling curve which demonstrates stepped (avalanche) behaviour (mcs = 45.7k – 47.2k). The respective Fourier transform of the lattice in the c-direction vs Monte Carlo steps (lower panel). **b,** The lattice structure shown at different

time points (i)-(iii). Occupied sites are drawn in white, and unoccupied sites are coloured according to the respective c-axis separation between occupied sites. **c,** Every c-direction gap between occupied sites are classified according to the distance, and the statistic is plotted as a population proportion vs time. 80 slices are taken from 0 to 80k mcs. The time points corresponding to **(b)** are marked with a dotted line (i),(ii),(iii). Results only shown up to 64k mcs ($\tilde{\mu} = 0$) for clarity.

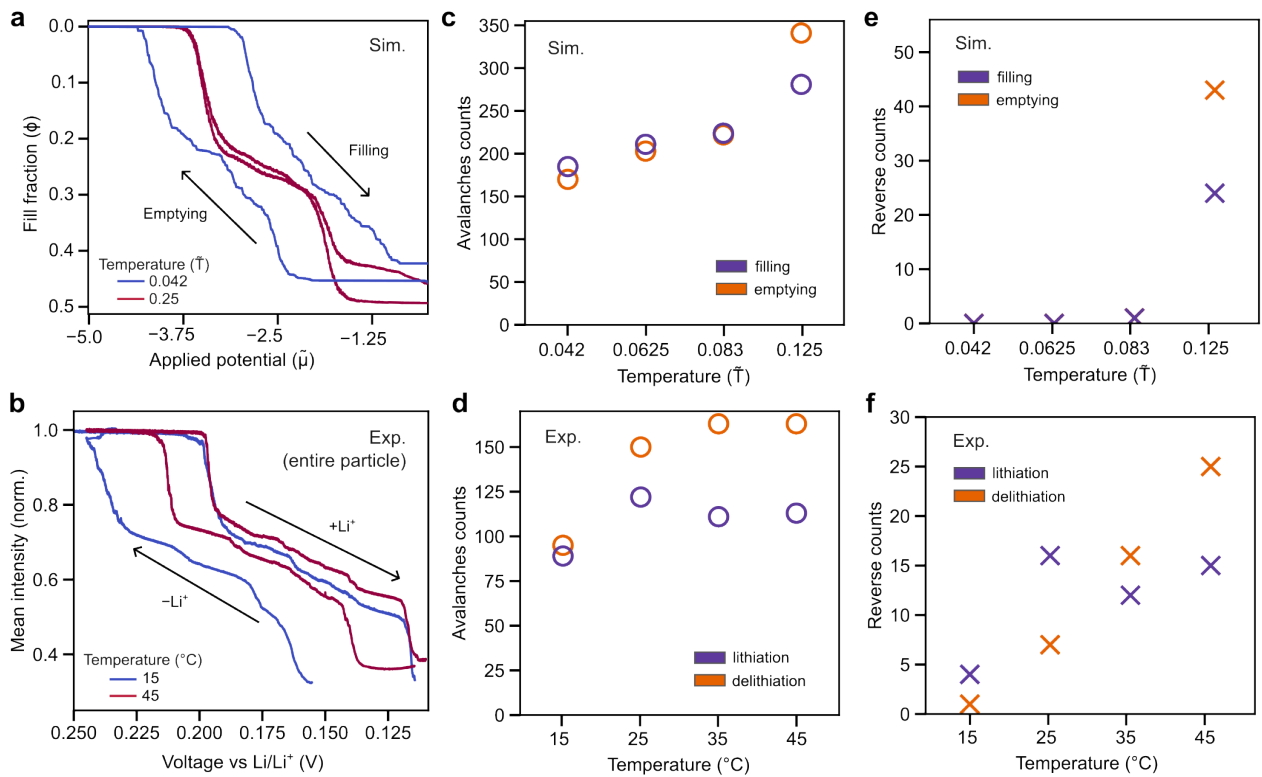

**Figure 4 | Temperature dependence of the filling dynamics and avalanche distributions. a,** Mean lattice fill fraction ($\phi$) vs applied potential ($\tilde{\mu}$) during filling and emptying simulations at different temperatures for the disordered model ($\tilde{h} = 0.4$, mcs = 80k) for a single run **b,** Mean optical intensity curves averaged over entire particle, versus voltage, normalised to the start of lithiation or end of delithiation. Lithiation and delithiation curves are both shown, direction denoted by black arrows. **c,d,** Detected number of avalanche events in model (**c**) and experiment (**d**) vs temperature. **e,f,** Detected number of reverse avalanche events in model (**e**) and experiment (**f**) vs temperature, where reverse avalanches are defined a step change in filling (**e**) or intensity (**f**) in the opposite direction of the driving force. The simulation avalanche counts are collected over a single run at each temperature. For the experimental counts, avalanches occurring in region A, B and C during the 1'L - 4L transition during de/lithiation in cycle 1 and 2 have been used.

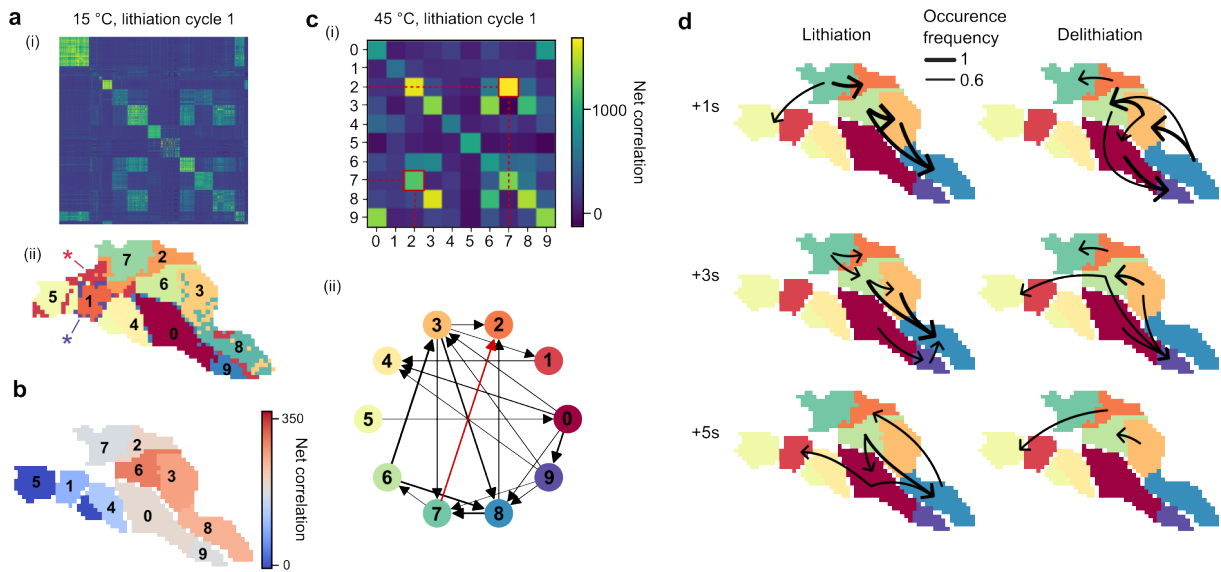

**Figure 5 | Spatio-temporal correlations during 1'L-4L transition. a**, Demonstration of concurrent activity mapping. Cross correlation is calculated from every pair of pixels (5 x 5 binned) on the particle, using the time-differential signal (dI/dt, cycle 1 at 15 °C). The resultant cross correlation function is summed for time-lags between -5s ≤ τ ≤ 5s, producing a symmetric cross correlation matrix which is clustered into 12 blocks using spectral co-clustering (i). These 12 clusters are then mapped back into space, which produces regions on the particle that show concurrent activity (ii). We exclude two clusters in further analyses, marked *, as they represent edge pixels. **b**, Identification of most and least connected regions. Correlation function from coarse-grained dataset (dI/dt, cycle 1,2 at 15-45 °C) is summed over time-lags (−5 s ≤ τ ≤ 5 s) and all inter-regions correlations are summed, excluding self-correlation, and mapped to the particle. A higher correlation (red) indicates the region experiences a large amount of correlation with other regions within ±5 s. **c**, Demonstration of intra-particle, positive time-lag correlations. Correlation function from coarse-grained dI/dt dataset (cycle 1 at 45 °C) is summed for positive time-lag (0.5s ≤ τ ≤ 3s) and the off-diagonal elements can be directly interpreted as successor / predecessor relationship between intercalation of two regions (i). This is mapped as a directed, weighted graph (ii), where each region (node) is connected by a directed edge with weight according to the difference of reciprocal values (i,j) - (j,i) from the 2D correlation matrix. The arrow hence points from the predecessor to the successor. For simplicity, 20 edges with the highest weights are shown. **d**, Time-lag dependent intra-particle correlations during de/lithiation. All predecessor / successor relationships are found for all de / lithiations (cycles 1 and 2 at 15, 25, 35, 45 °C) as per in **(c)**. From each graph, the most likely edges (predecessor / successor relationships) are found and mapped on to the particle at different lag times. The voltage window corresponding to the 1'L → 4L transition was used for all analyses.

## Methods

### *Sample preparation:*

Self-standing graphite electrode was prepared using a modified doctor blade technique. Briefly, polyvinylidene difluoride (PVDF HSV900, Kynar) was dissolved in N-methyl-2-pyrrolidone (NMP, Sigma-Aldrich). Conductive carbon powder (Super P Li, Timcal) and graphite powder (KS44 primary synthetic graphite, Timcal) was added to the solution and mixed into a slurry. The dry ratio of active material, conductive carbon and binder was 50:25:25 respectively. The slurry was cast onto a clean glass sheet using a doctor blade, dried at 80 °C and peeled off. The electrode was punched into 6 mm discs, dried overnight under vacuum at 120 °C and stored under argon. The final thickness of the electrode was ~40 μm.

### *Operando scattering microscopy:*

#### *Optical cell:*
A modified 2032 coin cell (shown in Supplementary Fig. S1a) was used with an optically accessible glass window (170 um, No 1.5H). The optical coin cell was assembled in a dry argon glovebox using lithium metal counter electrode, glass fibre separator (Whatman, GF/B) wet with LP57/VC2 electrolyte (1 M $LiPF_6$ in 3:7 wt. ratio of ethylene carbonate and ethyl methyl carbonate with 2% wt. of vinylene carbonate) and copper mesh current collector covering the active, self-standing electrode on the glass window side.

The coin cell was mounted in a custom-made temperature controlled holder (Supplementary Fig. S1b). The temperature of the holder was maintained using a PID controller (TED200C, Thorlabs) using a mounted peltier (TECF2S, Thorlabs) and temperature sensor (LM135Z). The temperature of the coin cell was calibrated with immersion oil using a PT100-sensor integrated optical coin cell.

#### *Optical set up*
The optical microscopy was adapted from our previous set up[24,25]. Briefly, the sample was epi-illuminated using an oil immersion objective (×100, 1.45NA, UPLXAPO, Olympus) by coupling a 4f (Kohler) illumination path from a fiber coupled LED (730 nm LED M730L5, M93L01, Thorlabs). The reflected and scattered light was imaged using an sCMOS camera (ORCA-Flash4.0 V3, Hamamatsu) with an overall magnification of ×167 and field of view of approximately 80 μm. The z-position of the sample was maintained within ±20 nm of the focal plane using the reflection profile of a 980 nm reference laser, as previously reported[24]. The final cell was mounted on a 3D piezo-controlled positioner (ECSx3030, Attocube) with a custom built stage and holder.

#### *Data acquisition:*

For the experiments presented in the main text, images were acquired at a frame rate of 2 Hz with 8 ms integration time over a smaller acquisition window (~ 500 x 450 pixels) using ImageJ [37]. Galvanostatic sequences were applied using a potentiostat (Gamry Interface 1010E). The cell was imaged at 25 °C during formation and 15, 25, 35 and 45 °C for later partial cycling.

*Image analysis:*
Details of analytical methods can be found in Supplementary methods.

### *Ex-situ charaterisation:*
The coin cell was disassembled and the self-standing electrode was extracted and washed in EC/EMC solvent and dried. SEM imaging was conducted using a TESCAN MIRA3 scanning electron microscope at 5.0 kV, with a 7.78 mm working distance using the in-beam secondary electron detector. Electron backscatter diffraction (EBSD) experiments were performed using a Thermo Fisher Scientific Helios 600i scanning electron microscope fitted with an Oxford Instruments Nordlys EBSD detector. The sample was mounted on a pre-tilted holder with additional stage tilt to achieve a total of 70°. Kikuchi patterns were recorded with an accelerating voltage of 20 kV and post-processed using Oxford Instruments AztecCrystal software.

### *Modelling:*
Modeling details and formulations can be found in Supplementary methods.

## Acknowledgements

The authors are especially grateful to S. Niblett for the insightful initial discussions, and thank A. Ashoka, R. Arul and H. Yang for the useful discussions and suggestions. J. H. acknowledges the Winton Programme for Physics of Sustainability for funding. G. S. P. acknowledges funding from a UK Engineering and Physical Sciences Research Council (EPSRC). A. J. M acknowledges financial support from Newnham College, Cambridge. C. P. G. acknowledges support from an ERC Advanced Investigator Grant (BATNMR, ERC 835073) and a Royal Society Research Professorship (180147). This work was supported by the Faraday Institution (FIRG060), a UKRI Frontier Research Grant (EP/Y015584/1) and by an Engineering Physical Sciences Research Council Programme Grant (EP/W017091/1). For the purpose of open access, the authors have applied a Creative Commons Attribution (CC BY) licence to any Author Accepted Manuscript version arising.

**Author contributions**

A.R. and C.P.G. conceived the idea and planned the experiments with J.H. and C.S; J.H. and J.L. prepared samples, with support from A.J.M.; J.H. constructed the optical set-up and temperature controlled cell with support from C.S.; J.H. performed the optical microscopy experiments and analyzed the data, with support from A.J.M and C.S.; R.L.J developed the model and the simulation code. J.H. performed the simulations under the guidance of R.L.J.; G.S.P. performed the SEM and EBSD measurements and analysed the data.; All authors discussed the results and contributed to writing the manuscript.

**Competing interests**

The authors declare no competing interests.